\documentclass[final,3p,times,authoryear]{elsarticle}
\usepackage{amssymb}
\usepackage{amsmath}
\usepackage{siunitx}

\journal{}

\begin{document}

\begin{frontmatter}


\title{Dislocation-based crystal plasticity simulation on grain-size dependence of mechanical properties in dual-phase steels}
\author{Misato Suzuki\fnref{label1}}
\author{Mayu Muramatsu\fnref{label2}}
\author{Kazuyuki Shizawa\corref{cor1}\fnref{label2}}
\ead{shizawa@mech.keio.ac.jp}
\cortext[cor1]{Corresponding author}
\affiliation[label1]{organization={Department of Science for Open and Environmental Systems, Graduate School of Keio University},
            addressline={3-14-1, Hiyoshi, Kohoku-ku},
            city={Yokohama},
            postcode={223-8522},
            state={Kanagawa},
            country={Japan}}
\affiliation[label2]{organization={Department of Mechanical Engineering, Keio University},
            addressline={3-14-1, Hiyoshi, Kohoku-ku},
            city={Yokohama},
            postcode={223-8522},
            state={Kanagawa},
            country={Japan}}
% \title{Dislocation-based crystal plasticity simulation on grain-size dependency of mechanical properties in dual-phase steels} %% Article title

% \author[label1]{Misato Suzuki}
% \author[label2]{Mayu Muramatsu}
% \author[label2]{Kazuyuki Shizawa}
% \affiliation[label1]{organization={Department of Science for Open and Environmental Systems, Graduate School of Keio University},
%             addressline={3-14-1, Hiyoshi, Kohoku-ku},
%             city={Yokohama},
%             postcode={223-8522},
%             state={Kanagawa},
%             country={Japan}}

% \affiliation[label2]{organization={Department of Mechanical Engineering, Keio University},
%             addressline={3-14-1, Hiyoshi, Kohoku-ku},
%             city={Yokohama},
%             postcode={223-8522},
%             state={Kanagawa},
%             country={Japan}}

% \author{} %% Author name

% \affiliation{organization={},%Department and Organization

%% Abstract
\begin{abstract}
In this study, the effect of ferrite grain size on the mechanical properties and dislocation behavior of dual-phase (DP) steel is investigated using dislocation-based crystal plasticity finite element analysis.
DP steel, composed of a soft ferritic phase and a hard martensitic phase, shows mechanical properties that are significantly influenced by ferrite grain size.
The mechanism underlying this grain size effect is clarified by analyzing the partitioning and distribution of stress, strain, and dislocations in each phase.
Three models with the same volume fraction of martensitic phase but different ferrite grain sizes are subjected to tensile loading.
Interestingly, even though only the ferrite grain size is changed, the stress in the martensitic phase exhibited a notable dependence on ferrite grain size.
This can be explained as follows.
Geometrically necessary (GN) dislocations accumulate on the ferrite side of the ferrite--martensite grain boundary, and the grain boundary occupancy per unit area increases as the ferrite grain size decreases.
As a result, smaller ferrite grain sizes make the ferritic phase less deformable owing to the effect of GN dislocations, shifting more deformation to the martensitic phase.   
This behavior is confirmed by the more uniform strain distribution and partitioning observed with decreasing ferrite grain size.
As the martensitic phase takes on greater deformation, the statistically stored dislocation density in the martensitic phase becomes ferrite grain size dependent, which in turn leads to the observed grain size dependence of stress in the martensitic phase.
\end{abstract}


% \begin{highlights}
% \item The effect of ferritic grain size on mechanical property of DP steel is clarified.
% \item The dislocation-based crystal plasticity finite element method is performed.
% \item Martensitic stress depends on ferritic grain size, even if martensitic size is fixed.
% \item Smaller grain size adds boundaries, dislocations increase near ferritic boundaries.
% \item As grain size decreases, ferrite deforms less, and martensite takes more deformation.
% \end{highlights}

%% Keywords
\begin{keyword}
Crystal plasticity finite element method \sep Dislocation \sep Dual-phase steel \sep Grain size effect


\end{keyword}

\end{frontmatter}



\section{Introduction}
\label{sec1}
To realize a low-carbon society, weight reduction of transportation equipment is an important issue.
In order to achieve high safety and good workability of equipment, there are high expectations for the development of structural materials with high strength and ductility.
Under these situations, the automobile industry is working on the application and further improvement of Advanced High Strength Steels (AHSS).
Dual-phase (DP) steel is a kind of AHSS that are widely used because of their combination of strength and ductility and low manufacturing cost.
DP steel consists of a soft phase (ferrite) and a hard phase (martensite), and a wide range of mechanical properties can be achieved by changing the ratio and distribution of the two phases.

One of the factors affecting the mechanical properties of DP steel is the volume fraction of martensitic phase $f_\mathrm{M}$ \citep{speich1979mechanical}.
Experiments have shown that $f_\mathrm{M}$ has an effect on strength such as yield stress and tensile strength \citep{peng1985effect,davies1978influence}, and ductility such as total elongation and uniform elongation \citep{kim1981effects,zhang2015effect}. 
The effect of $f_\mathrm{M}$ on the mechanical properties of DP steel has been reproduced not only by experiments but also by numerical calculations.
For example, experimentally based micromechanical analyses have shown that the $0.2\%$ proof stress depends on $f_\mathrm{M}$ \citep{liedl2002unexpected}.
The dependence of the tensile strength and the fracture strain on $f_\mathrm{M}$ has also been found to be owing to the change in the dominant damage mechanism as the volume fraction of martensite changes \citep{lai2015damage}.

The microstructures of DP steel are so complex that the mechanical properties are greatly influenced not only by $f_\mathrm{M}$, but also by the spatial distribution of each phase.
For example, even for the same $f_\mathrm{M}$, strength and ductility depend on whether the martensitic phase is fibrous or blocky \citep{zhang2015effect}, or whether it is networked or isolated (\citeauthor{park2014effect}, \citeyear{park2014effect}; \citeauthor{park2024quantitative}, \citeyear{park2024quantitative}).
It has also been shown that the mechanical properties of DP steel can vary even with the same morphology by altering the hardness of the martensitic phase through different heat treatment methods \citep{park2024effect}.

The effect of grain size on the mechanical properties of polycrystalline metals is one of the most important engineering topics for the development of high performance materials.
The increase in the strength of a material by decreasing the grain size is generally called the Hall-Petch effect \citep{hall1951deformation,petch1953cleavage}, and a grain size dependence of mechanical properties has been observed for a wide range of materials \citep{dini2010tensile,naghizadeh2019effects}.
In DP steel, the grain size of the ferritic phase has an effect on the yield  stress and tensile strength \citep{peng1985effect,davies1978influence}.
In addition, ultrafine-grained metals with grain size less than $1\,\si{\um}$ have been found to significantly improve strength \citep{valiev1993structure,tsuji2002strength}.
Ultrafine-grained DP steel have also been developed, and it is shown that strength is improved and ductility is not so much degraded when the ferrite grain size decreases \citep{calcagnotto2010effect}.
The behavior of ultrafine-grained metals has been reported to be realized by the fact that the mechanism governing plastic deformation varies with grain size \citep{punyafu2023microstructural}, but this is the case when focusing on single crystals and is not clear for polycrystals.

Since the development of high-performance materials and the clarification of deformation mechanisms through experiments as described above are very costly, it is important to predict mechanical properties from microstructures using numerical analysis.
The crystal plasticity theory \citep{asaro1977strain,asaro1979geometrical}, which models the simplified crystal slip as an elementary process and can consider the grain size and orientation of individual crystal grains, is often used in the analysis of metals \citep{takaki2010static,raabe2004using}.
To better reproduce the deformation of materials in which dislocations play an important role, dislocation-based crystal plasticity model have also been proposed \citep{ohashi1994numerical,zikry1996inelastic}.
In particular, the introduction of geometrically necessary (GN) dislocations makes it possible to describe the size effects of the stress--strain response \citep{evers2004scale}.

It is also important to consider dislocations in the crystal plasticity analysis of DP steels.
This is because the dislocations accumulated in the martensitic phase are much larger than those in the ferritic phase, which can have an effect on the mechanical properties.
% By considering dislocations in the crystal plasticity analysis of DP steel, the behavior of the cyclic loading \citep{eghtesad2020high,zecevic2016dual,kim2012crystal}, crystal orientation dependence \citep{jafari2019micromechanical}, and work hardening \citep{ramazani2013quantification} have been found to reproduce the experiments well.
By incorporating the effect of dislocations into the crystal plasticity analysis of DP steel, studies have successfully reproduced experimental results for cyclic loading behavior \citep{eghtesad2020high,zecevic2016dual,kim2012crystal} and work hardening \citep{ramazani2013quantification}, while \citet{jafari2019micromechanical} clarified that dislocation density in ferrite grains varies with crystal orientation, affecting the strength and ductility of DP steels.

The effect of grain size on materials has been analyzed using various methods such as molecular dynamics \citep{liu2013grain}, the phase-field method \citep{yeddu2018phase} and the finite element analysis introducing the effect of grain size into the composite law \citep{meyersm1982model}.
%%%ここのFEMのところだけ隣接する結晶粒の弾性異方性離散化手法の話でおかしい気がするがだとしてどう直したらいいかわからない．
%論文の内容としては，隣接する結晶粒の弾性異方性を考慮して複合則を改良？→塑性変形においても異方性を考慮して，粒径依存性を表すモデル化をしている．
In particular, the crystal plasticity analysis has been used to reproduce the grain size dependence of the mechanical properties of various metallic materials.
\citet{amelirad2019experimental} expressed the grain size dependence of stainless steel forming by adjusting the parameters of the hardening law according to grain size.
%ステンレス鋼
Based on the Hall-Petch relationship, methods of introducing grain size into the hardening law have also been proposed.
\citet{cruzado2018crystal} and \citet{lu2019dislocation} describe the grain size dependence of the mechanical properties of nickel alloys and nanograin copper, respectively, by introducing grain size into the hardening law.
\citet{lakshmanan2022crystal} also propose a model that can evaluate the effect of grain size and morphology of aluminum by using a representative length, which depends on the grain size and morphology, instead of a simple grain size.
Furthermore, \citet{ohashi2007multiscale} propose a grain-size dependent model in which the ease of emission of dislocation loops from a dislocation source in a grains is expressed as a function of grain size and incorporated into the critical resolved shear stress (CRSS).
\citet{ohno2007higher} propose a model that explains the grain size dependence of yield stress and the dislocation cell size dependence of flow stress by incorporating higher-order stresses arising from the self-energy of GN dislocations into the strain gradient theory.

The grain size dependence of DP steel has also been analyzed using crystal plasticity models. 
For example, a study by \citet{tasan2014strain} carries out a crystal plasticity analysis on the microstructure of DP steel obtained experimentally and shows that the mechanisms of strain localization and damage are affected by the grain size of the ferritic phase.
However, since the crystal plasticity analysis is based on experimentally obtained microstructures, it is difficult to separate the effect of ferrite grain size from the effect of the martensite morphology, and the effect of dislocations is not considered.
In a study by \citet{verma2016crystal} the dependence of the stress--strain relationship on the ferrite grain size is reproduced by constructing a crystal plasticity model that focuses on the effect of grain size.
However, the discussion on stress and strain partitioning for each phase is not sufficient.
\citet{woo2012stress} show that there is a dependence of mechanical properties on crystal orientation by considering stress and strain partitioning for each phase of DP steel, however, the effect of grain size has not been considered.
Thus, for DP steel, the effect of grain size on stress distribution, strain distribution, and mechanical properties of each phase has not been clarified based on dislocation behavior.

In this paper, the effect of ferrite grain size on the mechanical properties of DP steel is clarified using dislocation-based crystal plasticity analysis.
Specifically, several analytical models of DP steel with the same $f_\mathrm{M}$ and different ferrite grain sizes are prepared, and the stress--strain characteristics and spatial distributions of various quantities are computationally obtained when tensile load is applied to each model.
On the basis of the obtained results, the grain size dependence of the mechanical properties and the contributions of the microstructure consisting of ferritic and martensitic phases to the deformation are investigated and compared with the experimental results.
The grain size dependence of dislocation density and the dislocation density dependence of deformation response are also discussed.

\section{Dislocation-based crystal plasticity model}
\label{sec2}
The rate-type elastoviscoplastic constitutive equation of the crystal plasticity model used in this study is expressed as
\begin{equation}
    \overset{\triangledown}{\boldsymbol{T}}=\boldsymbol{C}^\mathrm{e}\colon\boldsymbol{D}^e=\boldsymbol{C}^\mathrm{e}:\boldsymbol{D}-\boldsymbol{C}^\mathrm{e}\colon\sum_{\alpha=1}^{N} \boldsymbol{P}^{(\alpha)}_\mathrm{S}\dot{\gamma}^{(\alpha)},
\label{eq1}
\end{equation}
where, $\overset{\triangledown}{\boldsymbol{T}}=\dot{\boldsymbol{T}}-\boldsymbol{W}^{*}\boldsymbol{T}+\boldsymbol{T}\boldsymbol{W}^{*}$ is the Mandel-Kratochvil rate of the Cauchy stress $\boldsymbol{T}$ with substructure spin $\boldsymbol{W}^{*}\equiv\boldsymbol{W}-\boldsymbol{W}^\mathrm{p}$.
The quantity $\dot{\boldsymbol{T}}$ denotes the material time derivative of the Cauchy stress, $\boldsymbol{W}$ is the continuum spin tensor, and $\boldsymbol{W}^\mathrm{p}$ is the plastic spin.
In addition, $\boldsymbol{C}^\mathrm{e}$ is the elastic modulus tensor, $\boldsymbol{D}$ is the deformation rate, $\dot{\gamma}^{(\alpha)}$ is the slip rate, and $\boldsymbol{P}^{(\alpha)}_\mathrm{S}=(\boldsymbol{s}^{(\alpha)}\otimes\boldsymbol{m}^{(\alpha)})_\mathrm{S}$ is the Schimid tensor.
The vectors $\boldsymbol{s}^{(\alpha)}$ and $\boldsymbol{m}^{(\alpha)}$ are unit vectors in the slip direction and normal to the slip plane, respectively, $(\bullet)_\mathrm{S}$ denotes the symmetric part of the second-order tensor, and $(\bullet)^{(\alpha)}$ denotes the value in the slip system $\alpha$.
Note that the exponential law proposed by \citet{hutchinson1976bounds} and \citet{pan1983rate} is used as the hardening law for the slip rate $\dot{\gamma}^{(\alpha)}$, as follows:
\begin{equation}
    \dot{\gamma}^{(\alpha)}=\dot{\gamma}^{(\alpha)}_{0}\mathrm{sgn}(\tau^{(\alpha)})\left\lvert\frac{\tau^{(\alpha)}}{g^{(\alpha)}}\right\rvert^{\frac{1}{m}},
    \label{eq2}
\end{equation}
where, $\mathrm{sgn}$ is the sign function, $\dot{\gamma}^{(\alpha)}_{0}$ is the reference slip rate, $g^{(\alpha)}$ is the flow stress, $m$ is the strain-rate sensitivity, and $\tau^{(\alpha)}$ is the resolved shear stress, expressed as $\tau^{(\alpha)}=\boldsymbol{T}\cdot\boldsymbol{P}^{(\alpha)}=\boldsymbol{s}^{(\alpha)}\cdot(\boldsymbol{T}\boldsymbol{m}^{(\alpha)})$.
The evolution equation of the flow stress in the crystal plasticity theory can be written as
\begin{equation}
    \dot{g}^{(\alpha)} = \sum_{\beta=1}^{N} h^{(\alpha\beta)} \left\lvert\dot{\gamma}^{(\beta)}\right\rvert,
    \label{eq3}
\end{equation}
where, $h^{(\alpha\beta)}$ denotes the hardening modulus that depends on the dislocation density.
To obtain the relation between hardening modulus and the dislocation density, we introduce the geometrically necessary (GN) dislocation density and the statistically stored (SS) dislocation density \citep{kujirai2020modelling}.
The rate of screw and edge components of GN dislocation density are expressed as \citep{kimura2020crystal}
\begin{equation}
    \dot{\rho}_{\mathrm{G,screw}}^{(\alpha)} =\frac{1}{\tilde{b}}\nabla\dot{\gamma}^{(\alpha)}\cdot\boldsymbol{t}^{(\alpha)},
    \label{eq4}
\end{equation}
\begin{equation}
    \dot{\rho}_{\mathrm{G,edge}}^{(\alpha)} =-\frac{1}{\tilde{b}}\nabla\dot{\gamma}^{(\alpha)}\cdot\boldsymbol{s}^{(\alpha)},
    \label{eq5}
\end{equation}
whrere, $\tilde{b}$ is the magnitude of Burgers vector, $\boldsymbol{t}^{(\alpha)}$ is the unit binormal vector defined by $\boldsymbol{t}^{(\alpha)}=\boldsymbol{s}^{(\alpha)}\times\boldsymbol{m}^{(\alpha)}$.
Thus, the net GN dislocation density is expressed as
\begin{equation}
    \rho_{\mathrm{G}}^{(\alpha)} =\sqrt{(\rho_{\mathrm{G,screw}}^{(\alpha)})^2+(\rho_{\mathrm{G,edge}}^{(\alpha)})^2},
    \label{eq6}
\end{equation}
where, $\rho_{\mathrm{G,screw}}$ and $\rho_{\mathrm{G,edge}}$ are the time integrated values of screw and edge components of GN dislocation density, respectively.
The evolution equation of the SS dislocation density are expressed as
\begin{equation}
    \dot{\rho}_{\mathrm{S}}^{(\alpha)} =\frac{c}{\tilde{b}L^{(\alpha)}}\left\lvert\dot{\gamma}^{(\alpha)}\right\rvert,
    \label{eq7}
\end{equation}
where $c$ is a numerical parameter on the order of $1$ and $L^{(\alpha)}$ is the dislocation mean free path \cite{ohashi1994numerical}.
The time integrated value of the SS dislocation density is expressd as
\begin{equation}
    \rho_{\mathrm{S}}^{(\alpha)} =\rho_{\mathrm{0}}^{(\alpha)}+\int\dot{\rho}_{\mathrm{S}}^{(\alpha)}\mathrm{d}t.
    \label{eq8}
\end{equation}
The relation between flow stress and dislocation density is known as the Bailey-Hirsch equation \citep{bailey1960dislocation}, which can be expressed in an extended form for a multiple-slip system as follows:
\begin{equation}
    g^{(\alpha)}= \tau_{y}^{(\alpha)}+a\mu\tilde{b}\sum_{\beta}\varOmega^{(\alpha\beta)}\sqrt{\rho^{(\beta)}_{\mathrm{h}}},
    \label{eq9}
\end{equation}
where, $\tau_{y}^{(\alpha)}$ is the lattice friction stress, $a$ is a numerical parameter on the order of $0.1$, $\mu$ is the shearing modulus, $\varOmega^{(\alpha\beta)}$ is the matrix representing the dislocation interaction between slip systems $\alpha$ and $\beta$, and $\rho^{(\beta)}_{\mathrm{h}}$ is the dislocation density contributing to work hardening, defined as $\rho^{(\beta)}_{\mathrm{h}}=\rho^{(\beta)}_{\mathrm{S}}$.
Eq.\eqref{eq9} does not include the third term, i.e., the grain size dependent term by \citet{ohashi2007multiscale}.
This is because we do not deal with ultrafine-grained materials in this study and the effect of the grain size dependent term can be assumed to be very small.
Then, comparing Eq.\eqref{eq3} with the time derivatives of Eq.\eqref{eq9}, we obtain the following relation between the hardening modulus and dislocation density.
\begin{equation}
    h^{(\alpha\beta)}= \frac{a\mu\tilde{b}\varOmega^{(\alpha\beta)}c}{2\tilde{b}L^{(\beta)}\sqrt{\rho^{(\beta)}_{\mathrm{h}}}}.
    \label{eq10}
\end{equation}
Note that the dislocation mean free path $L^{(\beta)}$ is expressed as a dislocation density-dependent form as follows \citep{ohashi1994numerical}:
\begin{equation}
    L^{(\beta)}= \frac{c^{*(\beta)}}{\sqrt{\sum_{\gamma} \omega^{(\beta\gamma)}\rho^{(\gamma)}_{\mathrm{L}}}},
    \label{eq11}
\end{equation}
where, $c^{*(\beta)}$ is the dislocation mobility and $\omega^{(\beta\gamma)}$ is the dislocation interaction matrix excluding the effect of self-hardening, and $\rho^{(\gamma)}_{\mathrm{L}}$ is the dislocation density contributing to the inhibition of the motion of mobile dislocations difined as $\rho^{(\gamma)}_{\mathrm{L}}=\rho^{(\gamma)}_{\mathrm{G}}+\rho^{(\gamma)}_{\mathrm{S}}$.

\section{Balance laws and FE analysis}
\label{sec3}
To analyze the deformation field using the finite element method (FEM), the balance law should be described in a weak form.
Here, we employ the rate form of the principle of virtual work in the form of updated Lagrangian, which can be used for large deformation problems.
Ignoring the body force, it is expressed as follows:
\begin{equation}
    \int_V\lbrack(\overset{\circ}{\boldsymbol{T}}-\boldsymbol{D}\boldsymbol{T}-\boldsymbol{T}\boldsymbol{D})\cdot\check{\boldsymbol{D}}+(\boldsymbol{L}\boldsymbol{T})\cdot\check{\boldsymbol{L}}\rbrack\mathrm{d}v=\oint_S\dot{\overset{(n)}{\boldsymbol{t}}}\cdot\check{\boldsymbol{v}}\mathrm{d}a,
    \label{eq12}
\end{equation}
where, $\overset{\circ}{\boldsymbol{T}}$ is the Jaumann rate of Cauchy stress, $\boldsymbol{L}$ is the velocity gradient, $\overset{(n)}{\boldsymbol{t}}$ is the traction, $\check{\boldsymbol{v}}$ is the virtual velocity, and $\check{(\bullet)}$ denotes the virtual value.
To introduce the rate form elastoviscoplastic constitutive equation of the crystal plasticity model used in this study into the principle of virtual work in Updated Lagrangian form, Eq.\eqref{eq1} is expressed in the Jaumann rate form as follows:
\begin{equation}
    \overset{\circ}{\boldsymbol{T}}=\boldsymbol{C}^\mathrm{e}:\boldsymbol{D}-\sum_{\alpha=1}^{N} (\boldsymbol{C}^\mathrm{e}\colon\boldsymbol{P}^{(\alpha)}_\mathrm{S}+\boldsymbol{P}^{(\alpha)}_\mathrm{A}\boldsymbol{T}-\boldsymbol{T}\boldsymbol{P}^{(\alpha)}_\mathrm{A})\dot{\gamma}^{(\alpha)},
\label{eq13}
\end{equation}
where, $(\bullet)_\mathrm{A}$ denotes the antisymmetric part of the second-order tensor, and the relation $\overset{\triangledown}{\boldsymbol{T}}=\overset{\circ}{\boldsymbol{T}}+\boldsymbol{W}^\mathrm{p}\boldsymbol{T}-\boldsymbol{T}\boldsymbol{W}^\mathrm{p}$ is used.
By substituting Eq.\eqref{eq13} into Eq.\eqref{eq12}, the finite element stiffness equations are derived.
The use of a viscoplastic-type hardening law requires a very small time step in the finite element analysis, which is not practical.
Therefore, in this study, the tangent modulus method \citep{peirce1984tangent} is employed so that the numerical analysis can be performed stably even for relatively large time increments.

\section{Results and discussions}
\label{sec4}
\begin{figure}[t]%% placement specifier
    \centering%% For centre alignment of image.
    \includegraphics[width=0.4\textwidth]{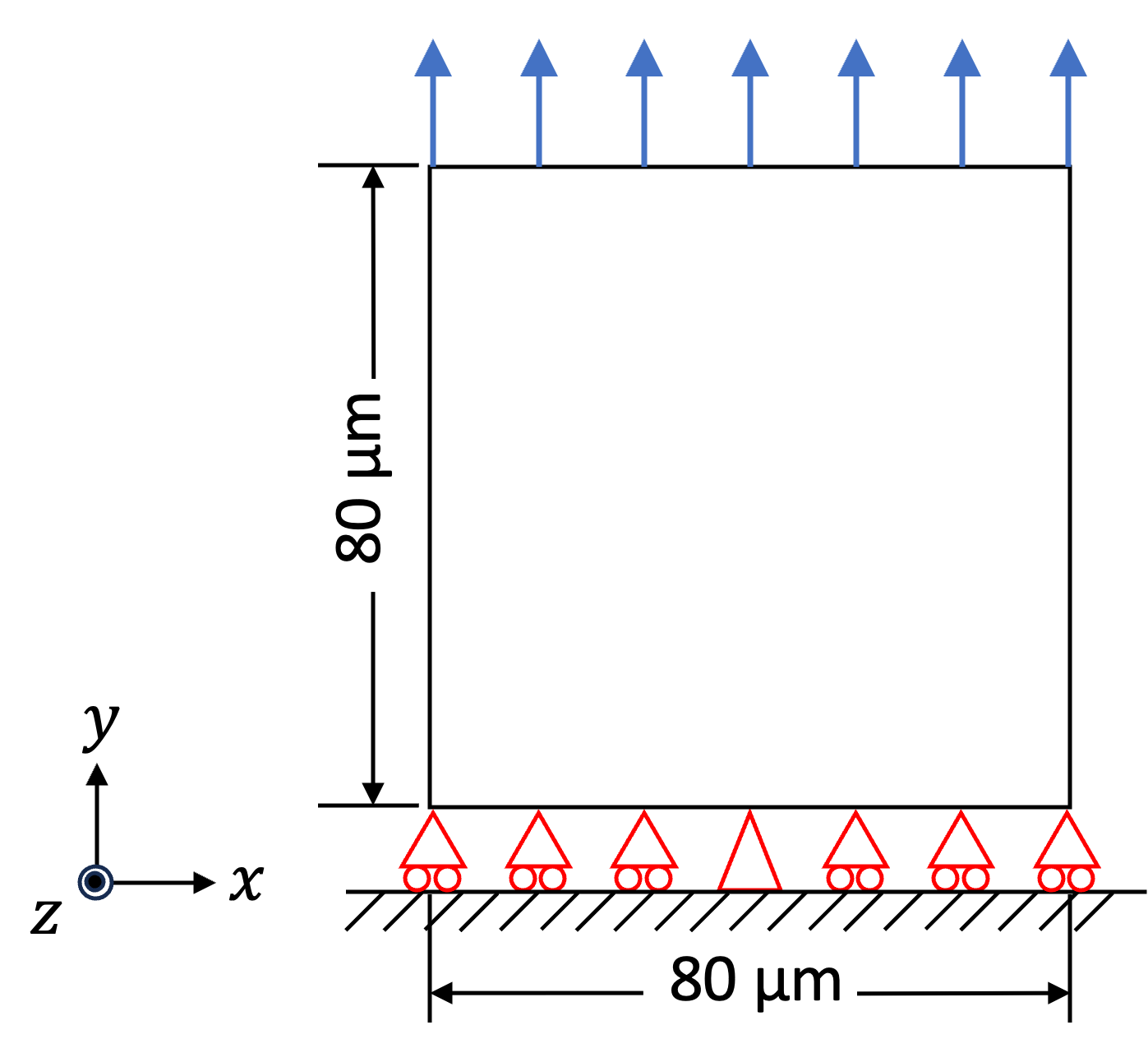}
    \caption{Analytical model}\label{fig1}
    \end{figure}
\begin{figure}[t]%% placement specifier
    \centering%% For centre alignment of image.
    \includegraphics[width=1\textwidth]{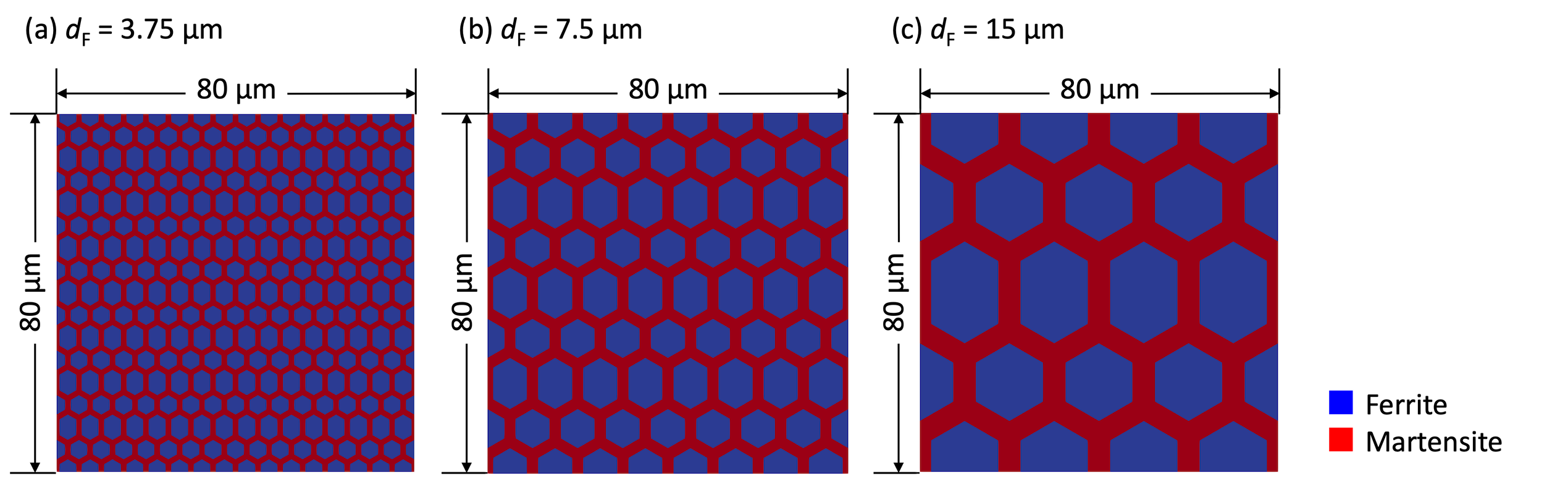}
    \caption{Phase distribution of analytical models. (a) Grain size of ferritic phase $d_\mathrm{F}=3.75\,\si{\um}$. (b) $d_\mathrm{F}=7.5\,\si{\um}$. (c) $d_\mathrm{F}=15\,\si{\um}$. The blue phase indicates the ferritic phase andthe red phase indicates the martensitic phase.}\label{fig2}
\end{figure}
\subsection{Analysis conditions}
\label{subsec1}
\begin{figure}[t]%% placement specifier
    \centering%% For centre alignment of image.
    \includegraphics[width=1\textwidth]{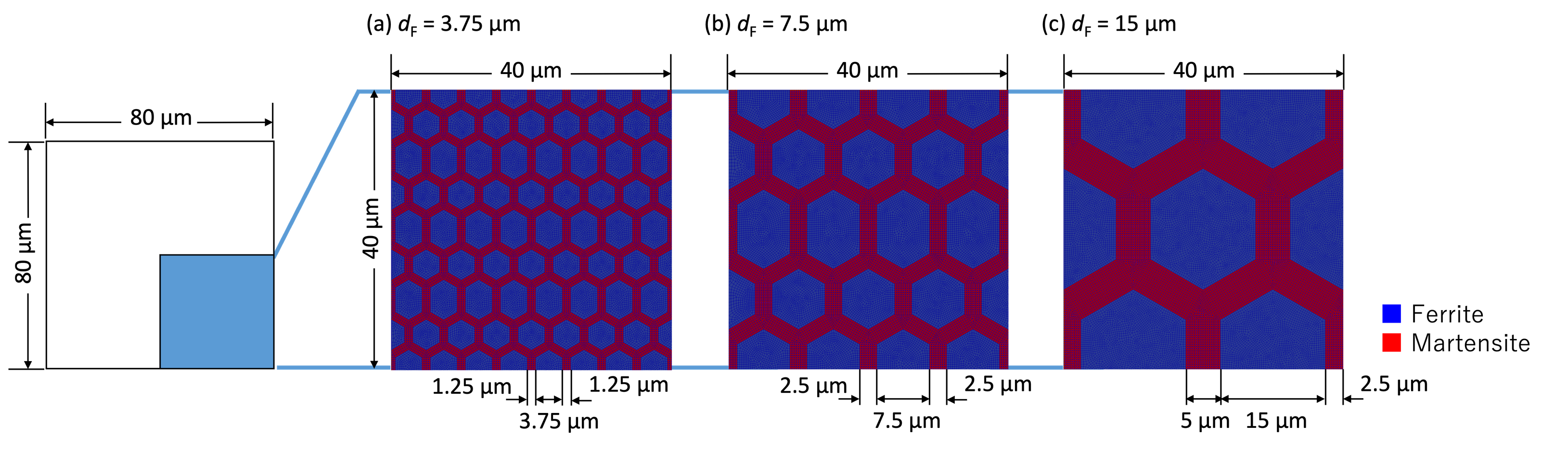}
    \caption{Detailed geometry of the analytical models. (a) Grain size of ferritic phase $d_\mathrm{F}=3.75\,\si{\um}$. (b) $d_\mathrm{F}=7.5\,\si{\um}$. (c) $d_\mathrm{F}=15\,\si{\um}$.}\label{fig3}
\end{figure}
\begin{figure}[t]%% placement specifier
    \centering%% For centre alignment of image.
    \includegraphics[width=1\textwidth]{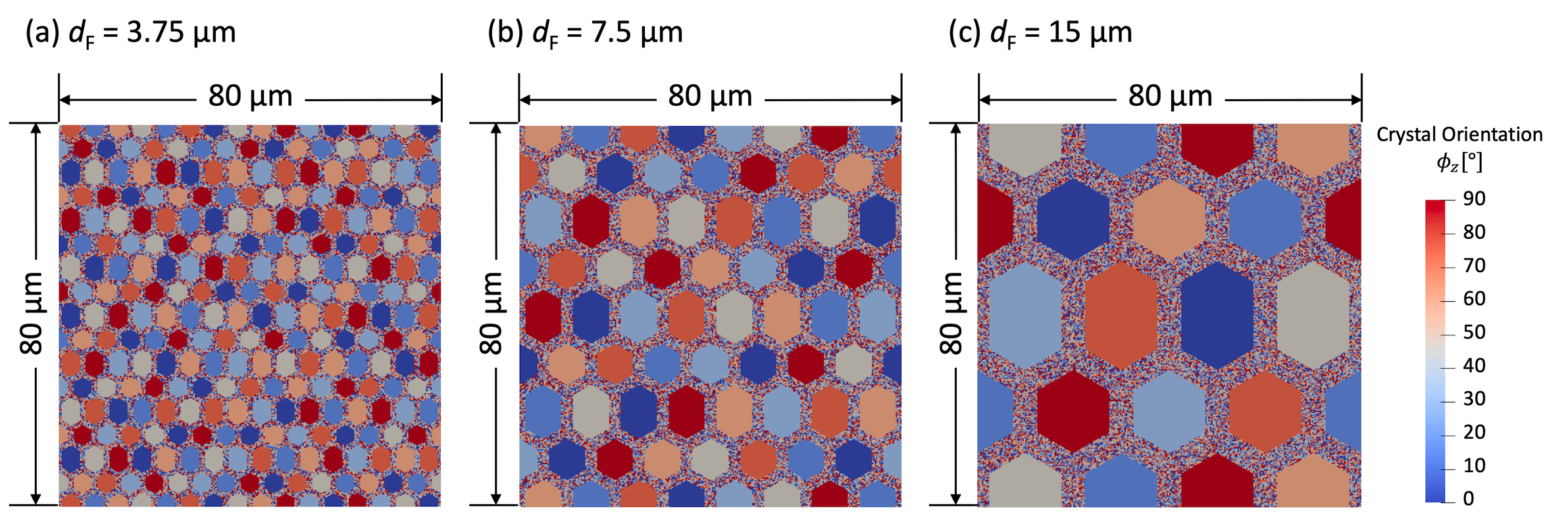}
    \caption{Distribution of crystal orientation in analytical models. (a) Grain size of ferritic phase $d_\mathrm{F}=3.75\,\si{\um}$. (b) $d_\mathrm{F}=7.5\,\si{\um}$. (c) $d_\mathrm{F}=15\,\si{\um}$.}\label{fig4}
\end{figure}
To investigate the effect of ferrite grain size on the mechanical properties of DP steel, the dislocation-based crystal plasticity FE analysis is performed.
The geometry of the specimen and boundary conditions in this analysis are shown in Fig.\ref{fig1}.
The analysis is performed on a square DP steel plate of $80\,\si{\um}\times80\,\si{\um}$.
A simple tensile displacement of $4\,\si{\um}$ toward the $y$-axis direction at the top surface of the specimen is applied and FE analysis is performed under plane strain conditions.
The time increment is set to $0.01\,\mathrm{s}$ and the strain rate to $0.1\,\mathrm{ks^{-1}}$.
Then, the microstructures to be analyzed are shown in Fig.\ref{fig2}.
The blue phase is the ferritic phase and the red phase is the martensitic phase, and one hexagon corresponds to one ferrite grain.
Three models [(a), (b) and (c)] with different grain size of the ferritic phase are prepared.
The area fractions of the martensitic phase are constant at $44\%$ for any grain size models.

To describe the geometry of the analytical models in detail, $40\,\si{\um}\times40\,\si{\um}$ region is extracted from the entire, $80\,\si{\um}\times80\,\si{\um}$ analytical domain and magnified in Fig.\ref{fig3}.
Figs.\ref{fig3}(a), (b) and (c) correspond to the magnified vies of Figs.\ref{fig2}(a), (b) and (c), respectively.
The ferrite grain sizes in Fig.\ref{fig3}(a), (b) and (c) are set to $d_\mathrm{F}=3.75$, $7.5$ and $15 \,\si{\um}$, respectively, while the martensite grain size is set constant at $d_\mathrm{M}=3.125 \,\si{\um}$ regardless of the difference in the ferrite grain size.
For martensite grains, each element corresponds to one grain [Fig.\ref{fig4}].

Distribution of crystal orientation in analytical models is shown in Fig.\ref{fig4}.
The fixed angles are used to express crystal orientation, and are expressed as $(\phi_x,\phi_y,\phi_z)$ when rotated in the order of $\phi_x$ around $x$ axis, $\phi_y$ around $y$ axis and, $\phi_z$ around $z$ axis for the fixed Cartesian coordinate system as shown in Fig.\ref{fig1}.
In this analysis, the crystal orientation of the ferritic and martensitic phases are set to $(\ang{0},\ang{0},\phi_z)$, and the value of $\phi_z$ is randomly selected from $\ang{0},\ang{15},\ang{30},\ang{45},\ang{60},\ang{75},\ang {90}$.
The material constants and numerical parameters used in this analysis are shown in Table \ref{tab1}.

\begin{table}[t]
\centering%% For centre alignment of tabular.
\caption{Material constants and numerical parameters.}\label{tab1}
\begin{tabular}{l l l}\hline
         &Ferrite &Martensite \\\hline
        Young's modulus $E$ & $205.9\,\mathrm{GPa}$ & $237.3\,\mathrm{GPa}$ \\
        Poisson's ratio $\nu$ & $0.3$ & $0.333$ \\
        Initial dislocation density $\rho_0$ & $1.0\,\si{\um^{-2}}$ &  $1000\,\si{\um^{-2}}$ \\
        Strain rate sensitivity $m$ & $0.007$ & $0.01$ \\ 
        Reference strain rate $\dot{\gamma}_0$ & $1.0\,\mathrm{ms^{-1}}$ &$1.0\,\mathrm{ms^{-1}}$\\
        Magnitude of burgers vector $\tilde{b}$& $0.249\,\mathrm{nm}$& $0.249\,\mathrm{nm}$\\
        $a$ & $0.1$ & $0.1$ \\
        $c$ & $1.1$ & $1.1$ \\
        $c^*\,\langle111\rangle\,\lbrace110\rbrace$ & $29$ & $10$\\
        $c^*\,\langle111\rangle\,\lbrace112\rbrace$ & $10$ & $10$\\
        Initial flow stress $g\,\langle111\rangle\,\lbrace110\rbrace$ & $24.9\,\mathrm{MPa}$ & $93.0\,\mathrm{MPa}$ \\
        Initial flow stress $g\,\langle111\rangle\,\lbrace112\rbrace$ & $29.9\,\mathrm{MPa}$ & $98.0\,\mathrm{MPa}$ \\
        Lattice friction constant $\tau_{y}\,\langle111\rangle\,\lbrace110\rbrace$& $23.0\,\mathrm{MPa}$ & $23.0\,\mathrm{MPa}$ \\
        Lattice friction constant $\tau_{y}\,\langle111\rangle\,\lbrace112\rbrace$& $28.0\,\mathrm{MPa}$ & $28.0\,\mathrm{MPa}$ \\ \hline
\end{tabular}
%% Use \caption command for table caption and label.
\end{table}

\subsection{Grain size dependence of stress and strain}
\label{subsec2}
\subsubsection{Stress--strain diagram}
\label{subsubsec1}
\begin{figure}[t]%% placement specifier
    \centering%% For centre alignment of image.
    \includegraphics[width=1\textwidth]{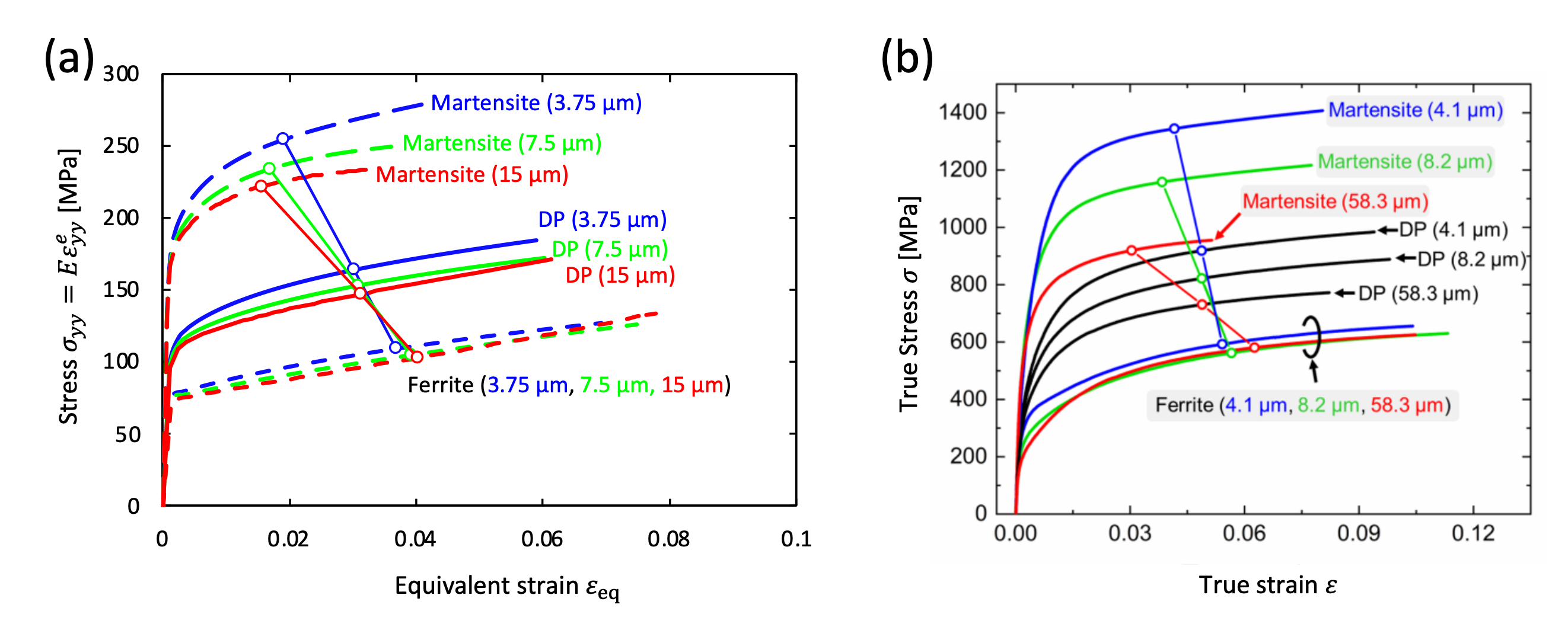}
    \caption{Stress--strain diagram. (a) Relationship between the stress $\sigma_{yy}$ and the equivalent strain $\varepsilon_{\mathrm{eq}}$ obtained by dislocation-based crystal plasticity analysis. (b) Stress--strain diagram obtained by the experiment \citep{myeong2017effect}.}\label{fig5}
\end{figure}

The results of dislocation-based crystal plasticity analysis for DP steel with three different ferrite grain sizes $d_\mathrm{F}$ under the analysis conditions shown in section \ref{subsec1} are presented.
First, to identify the role of each phase in contributing to the deformation, we base our analysis on the stress--strain diagram.
We employ the stress component $\sigma_{yy}$ as the value of the vertical axis in the stress--strain diagram so that the stress is equivalent to that in the experiment \citep{myeong2017effect}.
The stress component $\sigma_{yy}$ obtained from the Hooke's law by using elastic strain component $\varepsilon^{e}_{yy}$ under applied displacement toward the $y$ direction as follows:
\begin{equation}
    \sigma_{yy}=E\varepsilon^{e}_{yy}.
    \label{eq14}
\end{equation}
The relationship between the stress component $\sigma_{yy}$ and equivalent strain $\varepsilon_\mathrm{eq}$ is shown in Fig.\ref{fig5}(a).
The solid, dashed and long-dashed lines in Fig.\ref{fig5}(a) show the stress--strain curves for the entire DP steel, the ferritic phase and the martensitic phase, respectively.
Note that the stress components $\sigma_{yy}$ and equivalent strain $\varepsilon_\mathrm{eq}$ in the ferritic and martensitic phases $\bar{\sigma}_{yy}^\mathrm{F}$, $\bar{\sigma}_{yy}^\mathrm{M}$, $\bar{\varepsilon}_\mathrm{eq}^\mathrm{F}$ and $\bar{\varepsilon}_\mathrm{eq}^\mathrm{M}$ are obtained from the element averages of each phase.
Specifically, the total number of elements in the ferritic and martensitic phases is $n_\mathrm{F}$ and $n_\mathrm{M}$, respectively, and the average values of $\sigma_{yy}$ and $\varepsilon_\mathrm{eq}$ for each phase are calculated as follows:
\begin{equation}
    \bar{\sigma}_{yy}^\mathrm{F}=\frac{\sum_{i=1}^{n_\mathrm{F}}\sigma_{yy,i}}{n_\mathrm{F}},
    \quad\bar{\varepsilon}_\mathrm{eq}^\mathrm{F}=\frac{\sum_{i=1}^{n_\mathrm{F}}\varepsilon_{eq,i}}{n_\mathrm{F}},
    \label{eq15}
\end{equation}
\begin{equation}
    \bar{\sigma}_{yy}^\mathrm{M}=\frac{\sum_{i=1}^{n_\mathrm{M}}\sigma_{yy,i}}{n_\mathrm{M}},
    \quad\bar{\varepsilon}_\mathrm{eq}^\mathrm{M}=\frac{\sum_{i=1}^{n_\mathrm{M}}\varepsilon_{eq,i}}{n_\mathrm{M}}.
    \label{eq16}
\end{equation}
The stress--strain curves for DP steel, indicated by the solid line, tend to show higher stress with smaller ferrite grain size.
The stress--strain curves for the ferritic phase, indicated by the dashed line, remain almost the same regardless of the grain size, while the stress--strain curves for the martensitic phase, indicated by the long dashed line, remarkably show tendency for the stress to increase as the ferrite grain sizes decrease.
The reason for this tendency is discussed below in relation to dislocation behavior in section \ref{subsec3}.
Fig.\ref{fig5}(b) shows the stress--strain diagram obtained from the experiment by \citet{myeong2017effect}, which shows a similar trend to the stress--strain diagram obtained in this FE analysis.
The stress values on the vertical axis differ by about one order of magnitude between the analysis and the experiment.
This is because the analysis is based on microscopic specimens while the experiment is based on macroscopic bulk specimens.
To achieve quantitative agreement of stress values with experiments, a macroscopic analysis should be performed by applying the homogenization method to the FE analysis, which is the subject of our future work.
\begin{figure}[t]%% placement specifier
    \centering%% For centre alignment of image.
    \includegraphics[width=1\textwidth]{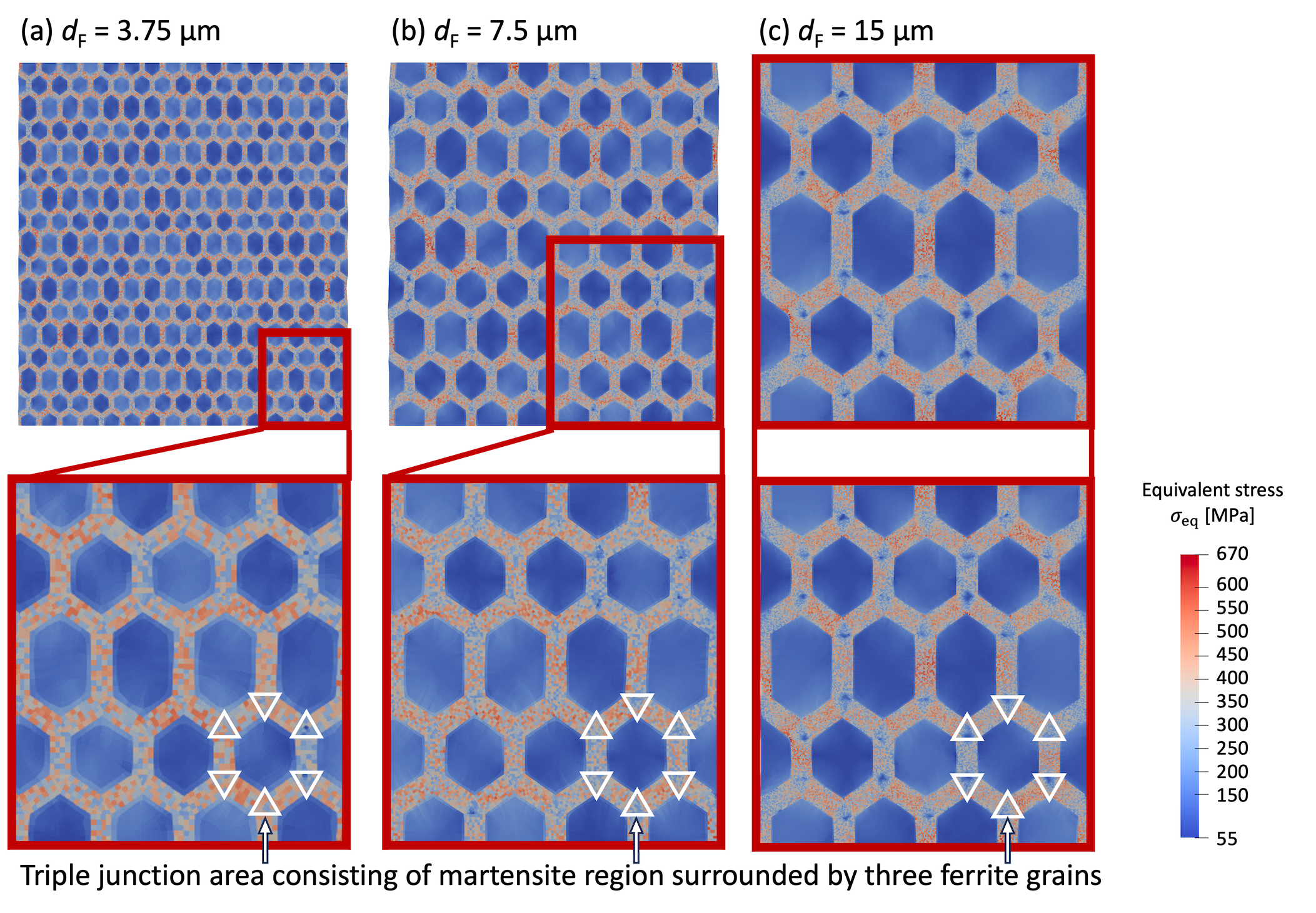}
    \caption{Equivalent stress distribution. (a) Grain size of ferritic phase $d_\mathrm{F}=3.75\,\si{\um}$. (b) $d_\mathrm{F}=7.5\,\si{\um}$. (c) $d_\mathrm{F}=15\,\si{\um}$. The upper figures show the distribution of the equivalent stress over the entire analytical domain. The lower figures show the results magnified so that the grain size of the ferritic phase is shown in the same size from (a) to (c). The white triangles highlight the triple junction area consisting of martensite region surrounded by the three ferrite grains.}\label{fig6}
\end{figure}
\clearpage
\begin{figure}[t]%% placement specifier
    \centering%% For centre alignment of image.
    \includegraphics[width=1\textwidth]{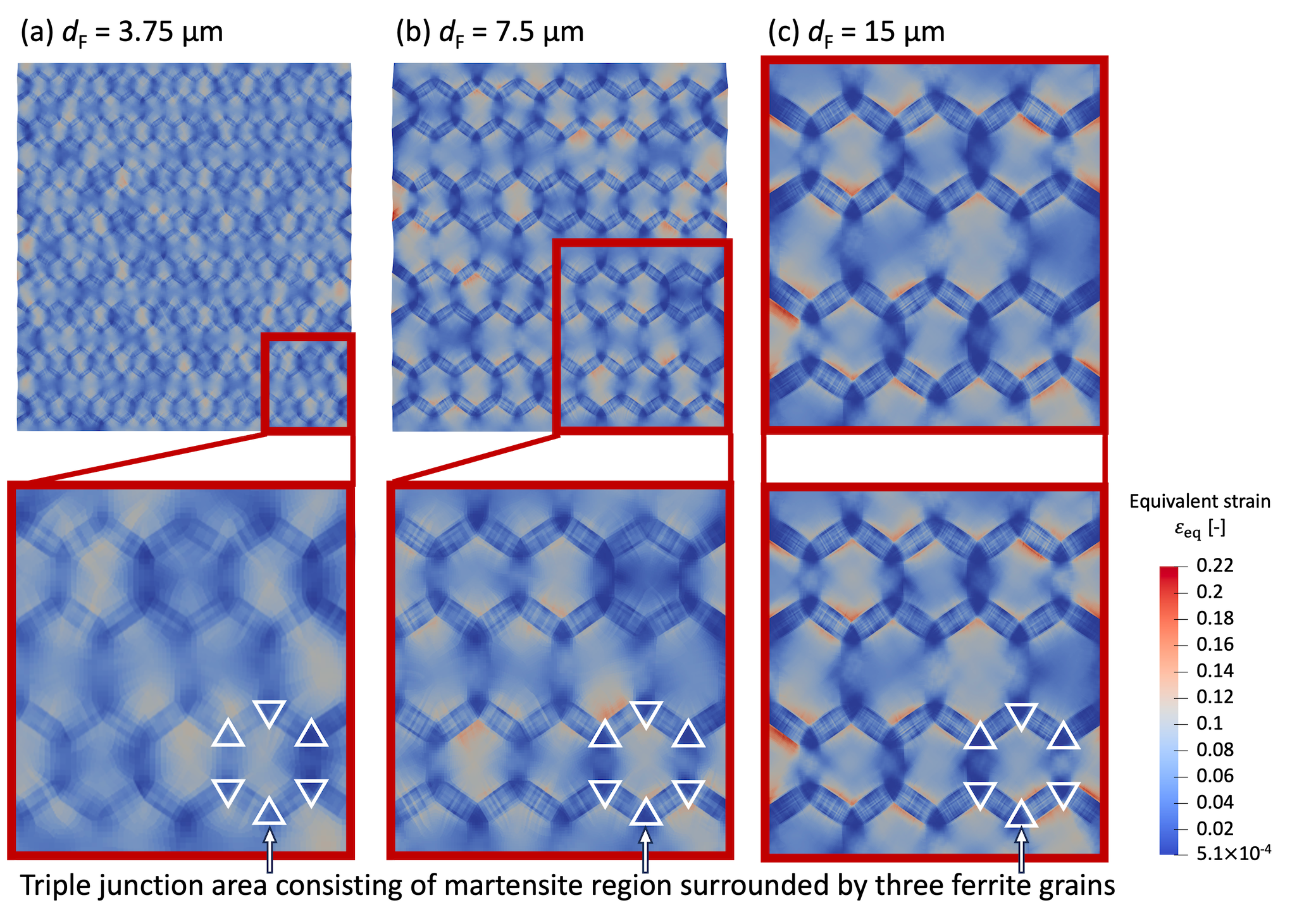}
    \caption{Equivalent strain distribution. (a) Grain size of ferritic phase $d_\mathrm{F}=3.75\,\si{\um}$. (b) $d_\mathrm{F}=7.5\,\si{\um}$. (c) $d_\mathrm{F}=15\,\si{\um}$. The upper figures show the distribution of the equivalent strain over the entire analytical domain. The lower figures show the results magnified so that the grain size of the ferritic phase is shown in the same size from (a) to (c). The white triangles highlight the triple junction area consisting of martensite region surrounded by the three ferrite grains.}\label{fig7}
\end{figure}

\subsubsection{Stress and strain distribution}
\label{subsubsec2}
In this section, the effects of ferrite grain size on the distributions of equivalent stress and equivalent strain are discussed.
Fig.\ref{fig6} shows the equivalent stress distribution.
Fig.\ref{fig6}(a), (b) and (c) are the results for $d_\mathrm{F}=3.75$, $7.5$ and $15 \,\si{\um}$, respectively.
The upper figures show the entire analytical domain, and the lower figures show the results magnified so that the grain size of the ferritic phase is shown in the same size from (a) to (c).
Such a magnified view is also applied to Figs.\ref{fig7}, \ref{fig11} and \ref{fig14}.
For all grain sizes $d_\mathrm{F}$, the martensitic phase shows the higher equivalent stress than ferritic phase.
Thus, the fact that hard phase such as the martensitic phase shows higher stress is consistent with the results by \citet{woo2012stress}, \citet{jafari2019micromechanical} and \citet{tasan2014strain}.
In Fig.\ref{fig6}, the white triangles highlight the triple junction area consisting of martensite region surrounded by the three ferrite grains.
In this highlighted zone, the equivalent stress is lower than in other regions of the martensitic phase.
This trend can be clearly seen in all such zones for $d_\mathrm{F}=15 \,\si{\um}$, however it becomes less clear as $d_\mathrm{F}$ decreases, and is not seen in most zones for $d_\mathrm{F}=3.75 \,\si{\um}$.
This is related to the more uniform strain distribution with decreasing grain size, as discussed below.

Fig.\ref{fig7} shows the equivalent strain distribution.
Fig.\ref{fig7}(a), (b) and (c) are the results for $d_\mathrm{F}=3.75$, $7.5$ and $15 \,\si{\um}$, respectively.
For all grain sizes $d_\mathrm{F}$, the ferritic phase shows the higher equivalent strain than the martensitic phase, and the difference in equivalent strain between the ferritic and martensitic phases depends on $d_\mathrm{F}$.
Thus, the fact that soft phase such as the ferritic phase shows higher strain is consistent with the results by \citet{tasan2014strain} and \citet{jafari2019micromechanical}.
In the case of $d_\mathrm{F}=15 \,\si{\um}$ [Fig.\ref{fig7}(c)], the martensitic phase contributes little to the deformation, and the ferritic phase shows large strain.
In Fig.\ref{fig7}, the white triangles highlight the triple junction area consisting of martensite region surrounded by the three ferrite grains.
In this highlighted zone, the equivalent strain is lower than in other regions of the martensitic phase.
The reason why this zone does not contribute to deformation in particular is that it has many martensite--martensite grain boundaries clustered away from the martensite--ferrite grain boundaries, which are more easily deformed.
In the case of $d_\mathrm{F}=7.5 \,\si{\um}$ [Fig.\ref{fig7}(b)], the equivalent strain in the martensitic phase is larger and the strain in the ferritic phase is compared to the case $d_\mathrm{F}=15 \,\si{\um}$.
As the grain size decreases further, that is, for $d_\mathrm{F}=3.75 \,\si{\um}$ [Fig.\ref{fig7}(a)], the equivalent strain in the martensitic phase becomes larger and that in the ferritic phase decreases more.
As for the martensite triple junction area, the smaller $d_\mathrm{F}$ is, the smaller the triple junction area itself becomes, and the fewer martensite--martensite grain boundaries assemble, resulting in a smaller difference between the equivalent strain in the triple junction area and in the other areas.
Thus, the smaller $d_\mathrm{F}$ is, the more the martensitic phase contributes to the deformation and the more uniform the equivalent strain is, which is discussed in detail in the next section \ref{subsubsec3}.

\subsubsection{Stress and strain partitioning}
\label{subsubsec3}
\begin{figure}[t]%% placement specifier
    \centering%% For centre alignment of image.
    \includegraphics[width=1\textwidth]{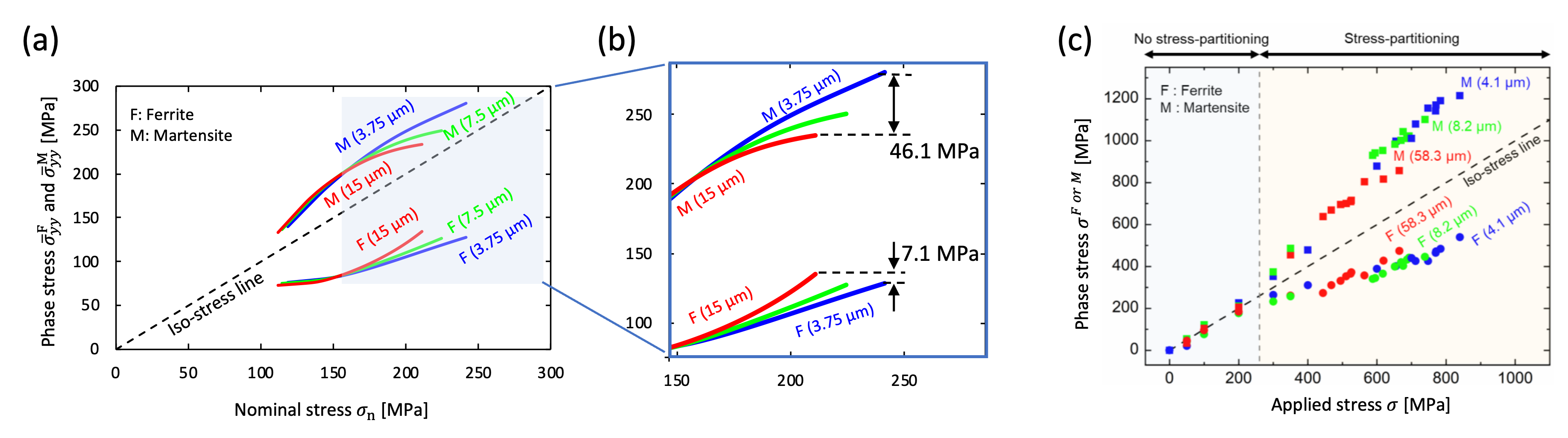}
    \caption{Stress partitioning. (a) Stress partitioning obtained by dislocation-based crystal plasticity analysis. (b) Magnified figure of (a). (c) Stress partitioning obtained by the experiment \citep{myeong2017effect}.}\label{fig8}
\end{figure}
\begin{figure}[t]%% placement specifier
    \centering%% For centre alignment of image.
    \includegraphics[width=1\textwidth]{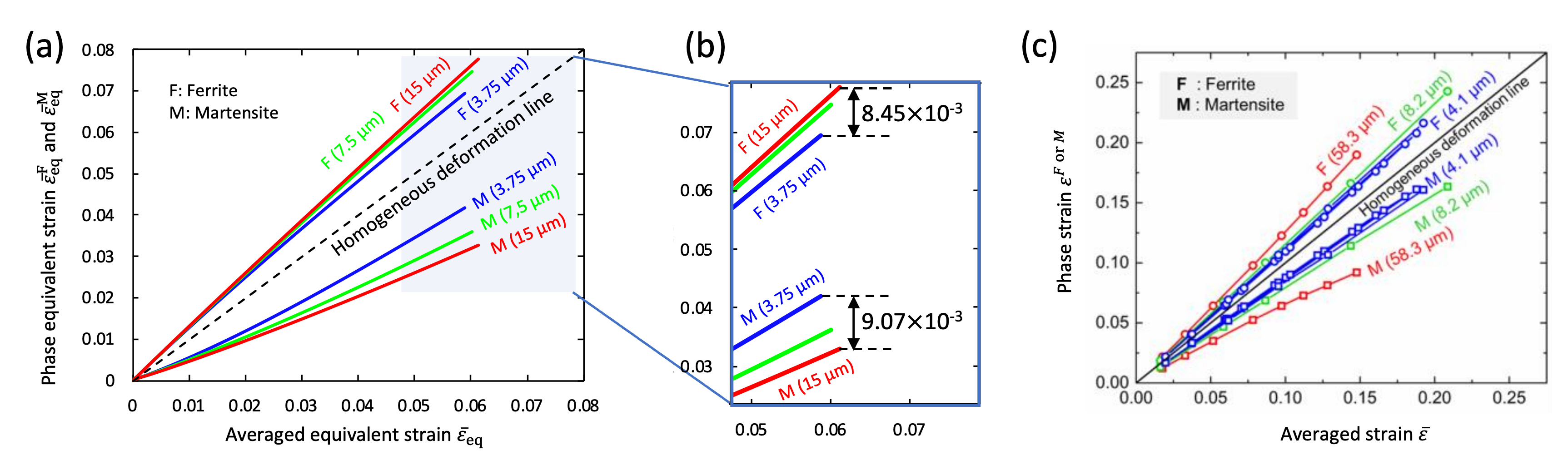}
    \caption{Strain partitioning. (a) Strain partitioning obtained by dislocation-based crystal plasticity analysis. (b) Magnified figure of (a) (c) Strain partitioning obtained by the experiment \citep{myeong2017effect}.}\label{fig9}
\end{figure}
The stress partitioning into those for ferritic and martensitic phases is shown in Fig.\ref{fig8}.
Fig.\ref{fig8}(a) shows the relationship of $\bar{\sigma}_{yy}^\mathrm{F}$ and $\bar{\sigma}_{yy}^\mathrm{M}$ in each phase to the nominal stress $\sigma_\mathrm{n}$, where the black dashed line is the iso-stress line.
The nominal stress $\sigma_\mathrm{n}$ is obtained by dividing the total $y$-directional components of the equivalent nodal force vector on the surface with applied displacement by the initial area of such surface.
On the other hand, the stress component $\sigma_{yy}$ is obtained using Eq.\eqref{eq14} from the elastic strain component $\varepsilon^{e}_{yy}$ and Hooke's law as described in section \ref{subsubsec1}.
The stress partitioning into the ferritic and martensitic phases, $\bar{\sigma}_{yy}^\mathrm{F}$ and $\bar{\sigma}_{yy}^\mathrm{M}$, is obtained by calculating the element averages for each phase from Eq.\eqref{eq15} and \eqref{eq16}, respectively.
For all grain sizes $d_\mathrm{F}$, $\bar{\sigma}_{yy}^\mathrm{M}$ is higher than $\bar{\sigma}_{yy}^\mathrm{F}$.
As the deformation proceeds and the nominal stress $\sigma_\mathrm{n}$ increases, the effect of ferrite grain size $d_\mathrm{F}$ on the stress increases.
Specifically, when the nominal stress $\sigma_\mathrm{n}$ is small, both $\bar{\sigma}_{yy}^\mathrm{M}$ and $\bar{\sigma}_{yy}^\mathrm{F}$ show little variation with $d_\mathrm{F}$. 
On the other hand, in case of high nominal stress $\sigma_\mathrm{n}$, the variation in $\bar{\sigma}_{yy}^\mathrm{M}$ with $d_\mathrm{F}$ becomes more pronounced than that in $\bar{\sigma}_{yy}^\mathrm{F}$.
Fig.\ref{fig8}(b) shows that the difference between the maximum $\bar{\sigma}_{yy}^\mathrm{M}$ at $d_\mathrm{F}=3.75 \,\si{\um}$ and the maximum $\bar{\sigma}_{yy}^\mathrm{M}$ at $d_\mathrm{F}=15 \,\si{\um}$ is greater than the corresponding difference in $\bar{\sigma}_{yy}^\mathrm{F}$.
That is, the change in $\bar{\sigma}_{yy}^\mathrm{M}$ with the change in $d_\mathrm{F}$ is larger than the change in $\bar{\sigma}_{yy}^\mathrm{F}$.
Fig.\ref{fig8}(c) shows the relationship between the stress for each phase and the nominal stress obtained in the experiment by \citet{myeong2017effect}.
By comparing Figs.\ref{fig8}(a) and (b) with (c), it can be said that the trend of stress partitioning obtained in this FE analysis is consistent with that of the experiment.

Then, the equivalent strain partitioning into ferritic and martensitic phases is shown in Fig.\ref{fig9}.
Fig.\ref{fig9}(a) shows the relationship of $\bar{\varepsilon}_\mathrm{eq}^\mathrm{F}$ and $\bar{\varepsilon}_\mathrm{eq}^\mathrm{M}$ in each phase to the averaged equivalent strain over the entire area of DP steel $\bar{\varepsilon}_{\mathrm{eq}}$, where the black dashed line is the homogeneous deformation line.
For all grain sizes $d_\mathrm{F}$, the ferritic phase shows the higher equivalent strain than martensitic phase.
As the deformation proceeds and the averaged equivalent strain $\bar{\varepsilon}_{\mathrm{eq}}$ increases, the effect of $d_\mathrm{F}$ on $\bar{\varepsilon}_\mathrm{eq}^\mathrm{F}$ and $\bar{\varepsilon}_\mathrm{eq}^\mathrm{M}$.
Specifically, when $\bar{\varepsilon}_{\mathrm{eq}}$ is small, the values of $\bar{\varepsilon}_\mathrm{eq}^\mathrm{F}$ and $\bar{\varepsilon}_\mathrm{eq}^\mathrm{M}$ is almost the same for each phase, regardless of $d_\mathrm{F}$.
On the other hand, in case of the high $\bar{\varepsilon}_{\mathrm{eq}}$, the smaller $d_\mathrm{F}$, the smaller the difference between $\bar{\varepsilon}_\mathrm{eq}^\mathrm{M}$ and $\bar{\varepsilon}_\mathrm{eq}^\mathrm{F}$, which indicates that the deformation becomes more uniform.
Fig.\ref{fig9}(b) shows that the difference between the maximum $\bar{\varepsilon}_\mathrm{eq}^\mathrm{M}$ at $d_\mathrm{F}=3.75 \,\si{\um}$ and the maximum $\bar{\varepsilon}_\mathrm{eq}^\mathrm{M}$ at $d_\mathrm{F}=15 \,\si{\um}$ is greater than the corresponding difference in $\bar{\varepsilon}_\mathrm{eq}^\mathrm{F}$.
That is, the change in $\bar{\varepsilon}_\mathrm{eq}^\mathrm{M}$ with the change in $d_\mathrm{F}$ is larger than the change in $\bar{\varepsilon}_\mathrm{eq}^\mathrm{F}$.
Fig.\ref{fig9}(c) shows the relationship between the equivalent strain for each phase and the averaged strain obtained in the experiment by \citet{myeong2017effect}.
By comparing Figs.\ref{fig9}(a) and (b) with (c), it can be said that the trend of strain partitioning obtained in this FE analysis is consistent with that of the experiment.

\subsection{Grain size dependence of dislocation density}
\label{subsec3}

\subsubsection{GN dislocation density}
\label{subsubsec4}
\begin{figure}[t]%% placement specifier
    \centering%% For centre alignment of image.
    \includegraphics[width=0.8\textwidth]{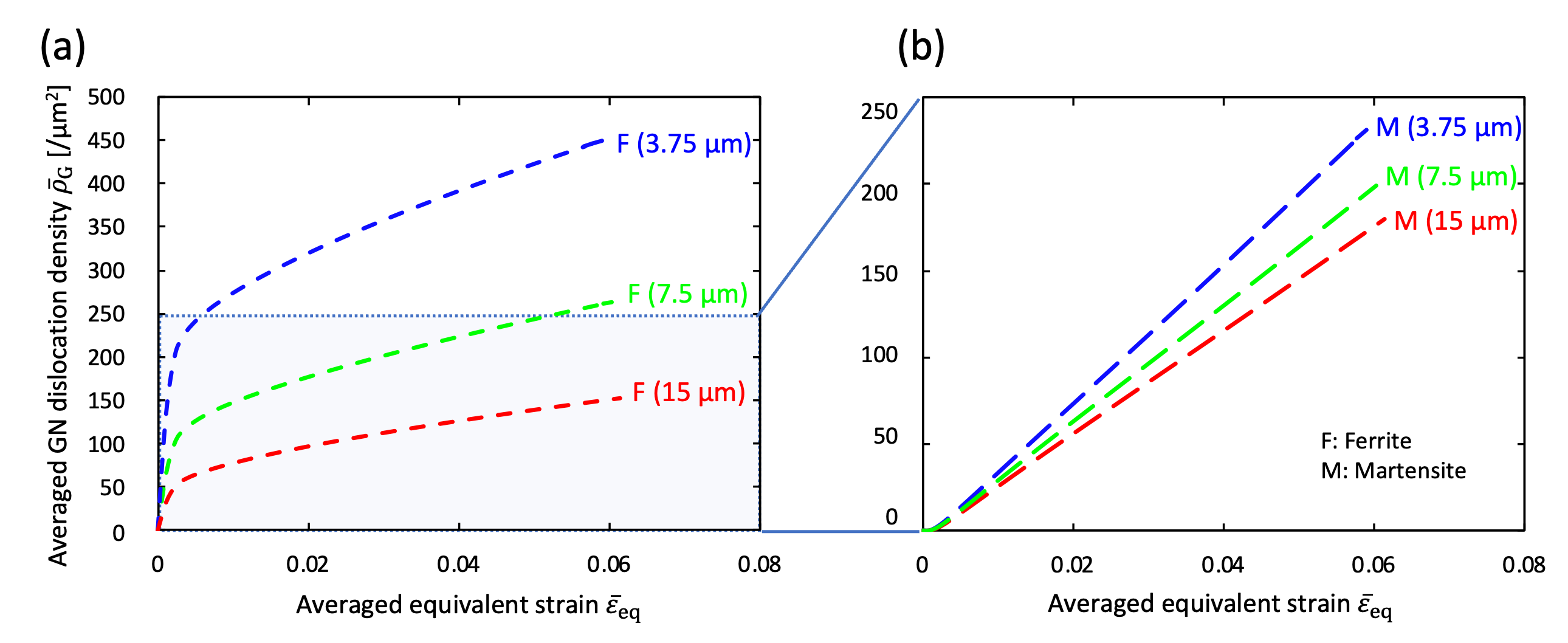}
    \caption{Relationship between averaged GN dislocation density and equivalent strain for each phase. (a) ferritic phase $\bar{\rho}_{\mathrm{G}}^\mathrm{F}$. (b) martensitic phase $\bar{\rho}_{\mathrm{G}}^\mathrm{M}$.}\label{fig10}
\end{figure}
\begin{figure}[t]%% placement specifier
    \centering%% For centre alignment of image.
    \includegraphics[width=1\textwidth]{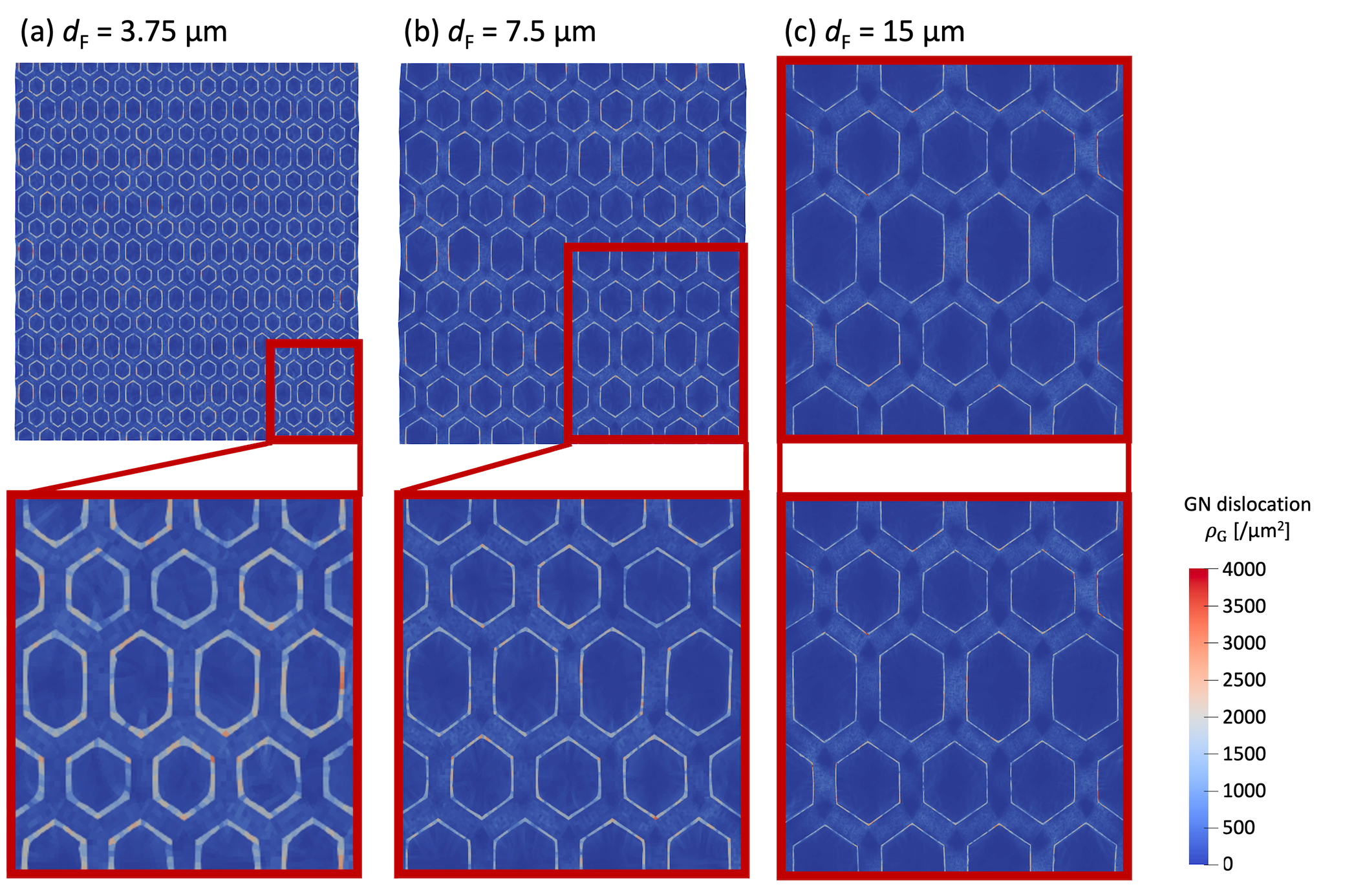}
    \caption{Distribution of GN dislocation density. (a) Grain size of ferritic phase $d_\mathrm{F}=3.75\,\si{\um}$. (b) $d_\mathrm{F}=7.5\,\si{\um}$. (c) $d_\mathrm{F}=15\,\si{\um}$. The upper figures show the distribution of GN dislocation density over the entire analytical domain. The lower figures show the results magnified so that the grain size of the ferritic phase is shown in the same size from (a) to (c).}\label{fig11}
\end{figure}
In this section, the effect of the ferrite grain size on the GN dislocation density $\rho_{\mathrm{G}}$ is discussed.å
Fig.\ref{fig10} shows the relationship between the averaged equivalent strain $\bar{\varepsilon}_{\mathrm{eq}}$ of the entire DP steel and the averaged GN dislocation density of the ferritic phase $\bar{\rho}_{\mathrm{G}}^\mathrm{F}$ [Fig.\ref{fig10}(a)] and that of the martensitic phase $\bar{\rho}_{\mathrm{G}}^\mathrm{M}$ [Fig.\ref{fig10}(b)].
For both phases, the averaged GN dislocation density becomes larger for smaller grain size $d_\mathrm{F}$.

Fig.\ref{fig10}(a) shows that $\bar{\rho}_{\mathrm{G}}^\mathrm{F}$ increases rapidly in the early stage of deformation and then increases more slowly as deformation proceeds.
This is because the first-order gradient of slip rate at the ferrite--martensite grain boundary is very large in the early stage of deformation, and GN dislocations accumulate immediatly on the ferrite side of the ferrite--martensite grain boundary.
The reason why the rate of this increase differs with $d_\mathrm{F}$ can be seen from the distribution of GN dislocation density shown in Fig.\ref{fig11}.
Fig.\ref{fig11}(a), (b) and (c) show the distribution of GN dislocation density for $d_\mathrm{F}=3.75$, $7.5$ and $15 \,\si{\um}$, respectively.
For all grain sizes, the GN dislocation density is larger on the ferrite side of the ferrite--martensite grain boundary, and there is no significant difference in magnitude with grain size.
Nevertheless, the reason why the averaged GN dislocation density is higher in the ferritic phase for smaller $d_\mathrm{F}$ in Fig.\ref{fig10}(a) is because the region on the ferrite side of the ferrite--martensite grain boundary increases for smaller $d_\mathrm{F}$.
Specifically, the length of the ferrite--martensite grain boundary per unit area ($1\,\si{\um^2}$) are $0.573$, $0.287$ and $0.143 \,\si{\um}$ for $d_\mathrm{F}=3.75$, $7.5$ and $15 \,\si{\um}$, respectively, which indicates that the smaller $d_\mathrm{F}$ is, the longer the grain boundary per unit area is.
This means that the smaller $d_\mathrm{F}$ is, the less bulk area in the ferritic phase contributes to deformation.
Therefore, when specimens for different grain sizes are under the same applied displacement, the smaller $d_\mathrm{F}$ is, the more the martensitic phase has to contribute to the deformation.
Here, the GN dislocation density of the martensitic phase $\bar{\rho}_{\mathrm{G}}^\mathrm{M}$ in Fig.\ref{fig10}(b) shows a dependence of the GN dislocation density on the ferrite grain size, even though the grain size of the martensite is constant.
This indicates that the smaller $d_\mathrm{F}$ causes the lower deformation ability of the ferritic phase, which requires deformation in the martensitic phase, and thus the more GN dislocations accumulate at the grain boundary in the martensitic phase.

\begin{figure}[t]%% placement specifier
    \centering%% For centre alignment of image.
    \includegraphics[width=1\textwidth]{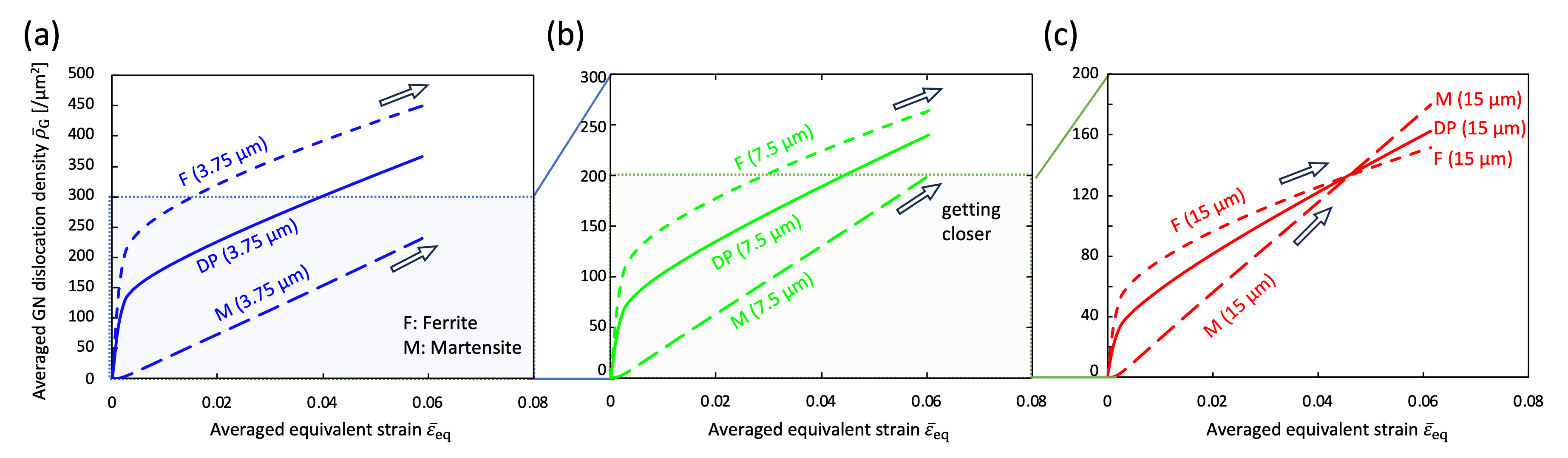}
    \caption{Relationship between averaged GN dislocation density and equivalent strain for different ferrite grain sizes. (a) Grain size of ferritic phase $d_\mathrm{F}=3.75\,\si{\um}$. (b) $d_\mathrm{F}=7.5\,\si{\um}$. (c) $d_\mathrm{F}=15\,\si{\um}$.}\label{fig12}
\end{figure}

The GN dislocation densities of the ferritic and martensitic phases are then compared for each ferrite grain size.
Fig.\ref{fig12} shows the relationship between the averaged equivalent strain $\bar{\varepsilon}_{\mathrm{eq}}$ over the entire DP steel and averaged GN dislocation density $\bar{\rho}_{\mathrm{G}}$ for each grain size.
In the case of $d_\mathrm{F}=3.75 \,\si{\um}$ [Fig.\ref{fig12}(a)] and $d_\mathrm{F}=7.5 \,\si{\um}$ [Fig.\ref{fig12}(b)], averaged GN dislocation densities in the ferritic phase $\bar{\rho}_{\mathrm{G}}^\mathrm{F}$ are larger than that in the martensitic phase $\bar{\rho}_{\mathrm{G}}^\mathrm{M}$ throughout whole deformation.
On the other hand, with $d_\mathrm{F}=15 \,\si{\um}$ [Fig. \ref{fig12}(c)], $\bar{\rho}_{\mathrm{G}}^\mathrm{F}$ is larger than $\bar{\rho}_{\mathrm{G}}^\mathrm{M}$ in the early stage of deformation, however, the relationship reverses in the middle stage and $\bar{\rho}_{\mathrm{G}}^\mathrm{M}$ becomes larger than $\bar{\rho}_{\mathrm{G}}^\mathrm{F}$.
Fig.\ref{fig12}(b) for $d_\mathrm{F}=7.5 \,\si{\um}$ shows that the difference between $\bar{\rho}_{\mathrm{G}}^\mathrm{F}$ and $\bar{\rho}_{\mathrm{G}}^\mathrm{M}$ is gradually decreasing.
The same reversal is expected to occur as $d_\mathrm{F}=15 \,\si{\um}$ when the deformation is further proceeded even in the case of small grain size.
The inversion of GN dislocation density occurs because GN dislocations on the ferrite side of the ferrite--martensite grain boundary become less likely to accumulate as deformation proceeds, while in the martensitic phase, GN dislocations continue to accumulate at a constantly owing to the large amount of martensite--martensite grain boundaries.
In addition, the inversion occurs at a earlier stage of deformation for larger $d_\mathrm{F}$, because the larger $d_\mathrm{F}$ is, the smaller the length of ferrite--martensite grain boundary is.
The larger the bulk region of the ferritic phase is, the smaller the GN dislocation density in the ferritic phase [Fig.\ref{fig10}(a)], and the rate of increase in GN dislocation density becomes also smaller.
% In addition, the inversion occurs at an earlier stage of deformation with a larger ferrite grain size $d_\mathrm{F}$ because there are fewer ferrite-martensite grain boundaries with a larger $d_\mathrm{F}$.
% The larger $d_\mathrm{F}$ means that the ferrite--martensite grain boundary is fewer and the bulk region of the ferritic phase is large.
% This make the GN dislocation density in the ferritic phase small [Fig.\ref{fig10}(a)] and rate of increase also small.
%In addition,から始まる一文が長すぎだと思った時のために上に切り分けた文章を残しておく

\subsubsection{SS dislocation density}
\label{subsubsec5}
\begin{figure}[t]%% placement specifier
    \centering%% For centre alignment of image.
    \includegraphics[width=0.45\textwidth]{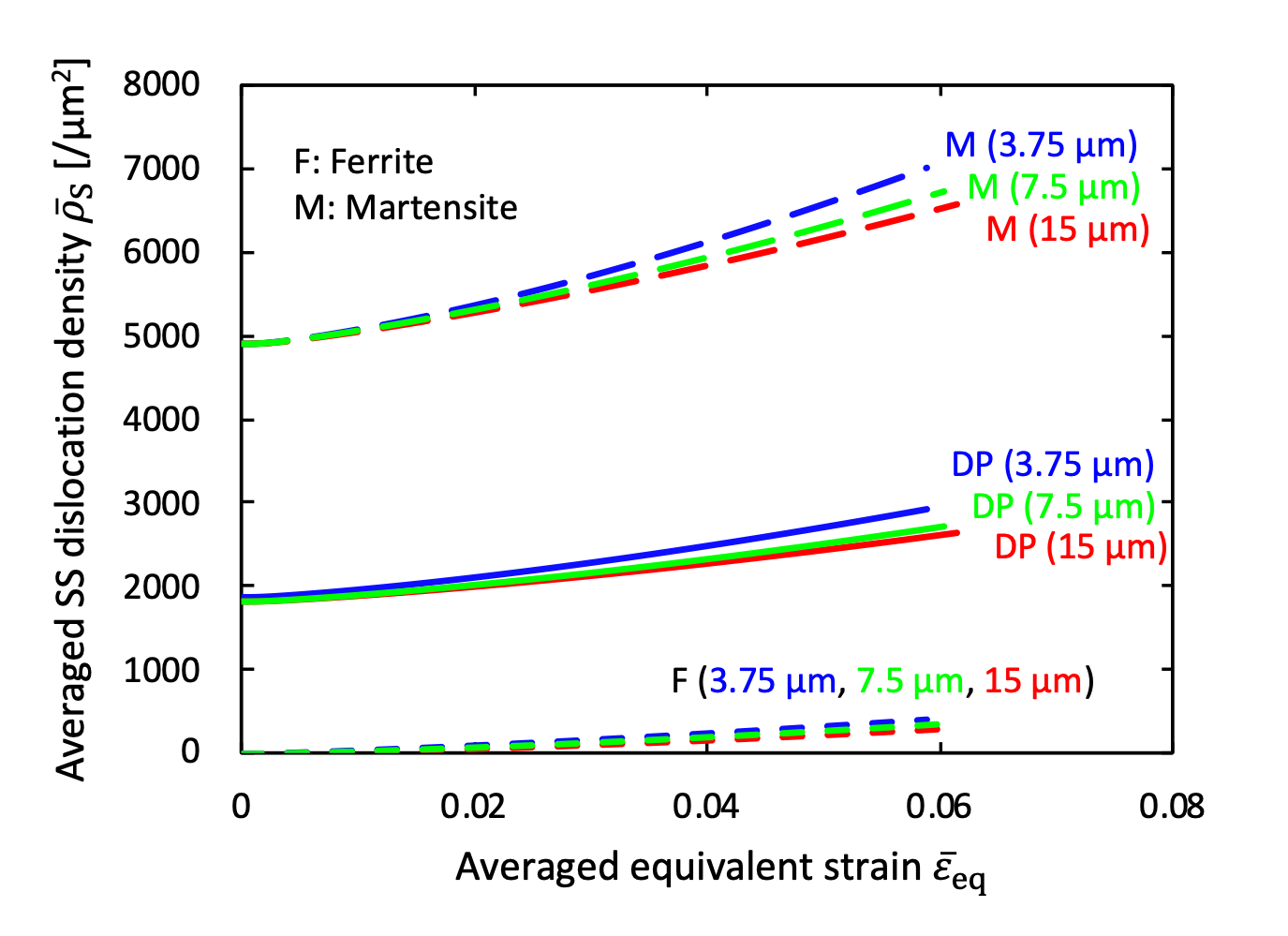}
    \caption{Relationship between averaged SS dislocation density and equivalent strain.}\label{fig13}
\end{figure}

In this section, the effect of different ferrite grain size on SS dislocation density $\rho_{\mathrm{S}}$ is discussed.
Fig.\ref{fig13} shows the relationship between the averaged equivalent strain $\bar{\varepsilon}_{\mathrm{eq}}$ and the averaged SS dislocation density $\bar{\rho}_{\mathrm{S}}$ in the ferritic phase, martensitic phase and entire DP steel.
The averaged SS dislocation density in the martensitic phase $\bar{\rho}_{\mathrm{S}}^\mathrm{M}$ is larger that in the ferritic phase $\bar{\rho}_{\mathrm{S}}^\mathrm{F}$ from the beginning, and this trend does not change throughout the deformation.
This is because the martensitic phase has a large initial SS dislocation density owing to lattice-invariant shear at its generation.
Both $\bar{\rho}_{\mathrm{S}}^\mathrm{F}$ and $\bar{\rho}_{\mathrm{S}}^\mathrm{M}$ are larger as $d_\mathrm{F}$ is smaller, and the difference in $\bar{\rho}_{\mathrm{S}}^\mathrm{M}$ for different $d_\mathrm{F}$ is more significant than that in $\bar{\rho}_{\mathrm{S}}^\mathrm{F}$.
The grain size dependence of SS dislocation density in the martensitic phase can be understood from Eq.\eqref{eq7} and Eq.\eqref{eq11} through the fact that the smaller  $d_\mathrm{F}$, the more the martensitic phase is taking on deformation, and the higher the slip rate in the martensitic phase is.

As a result, the grain size dependence of work hardening appears in the martensitic phase as shown in Fig.\ref{fig5}(a).

\begin{figure}[t]%% placement specifier
    \centering%% For centre alignment of image.
    \includegraphics[width=1\textwidth]{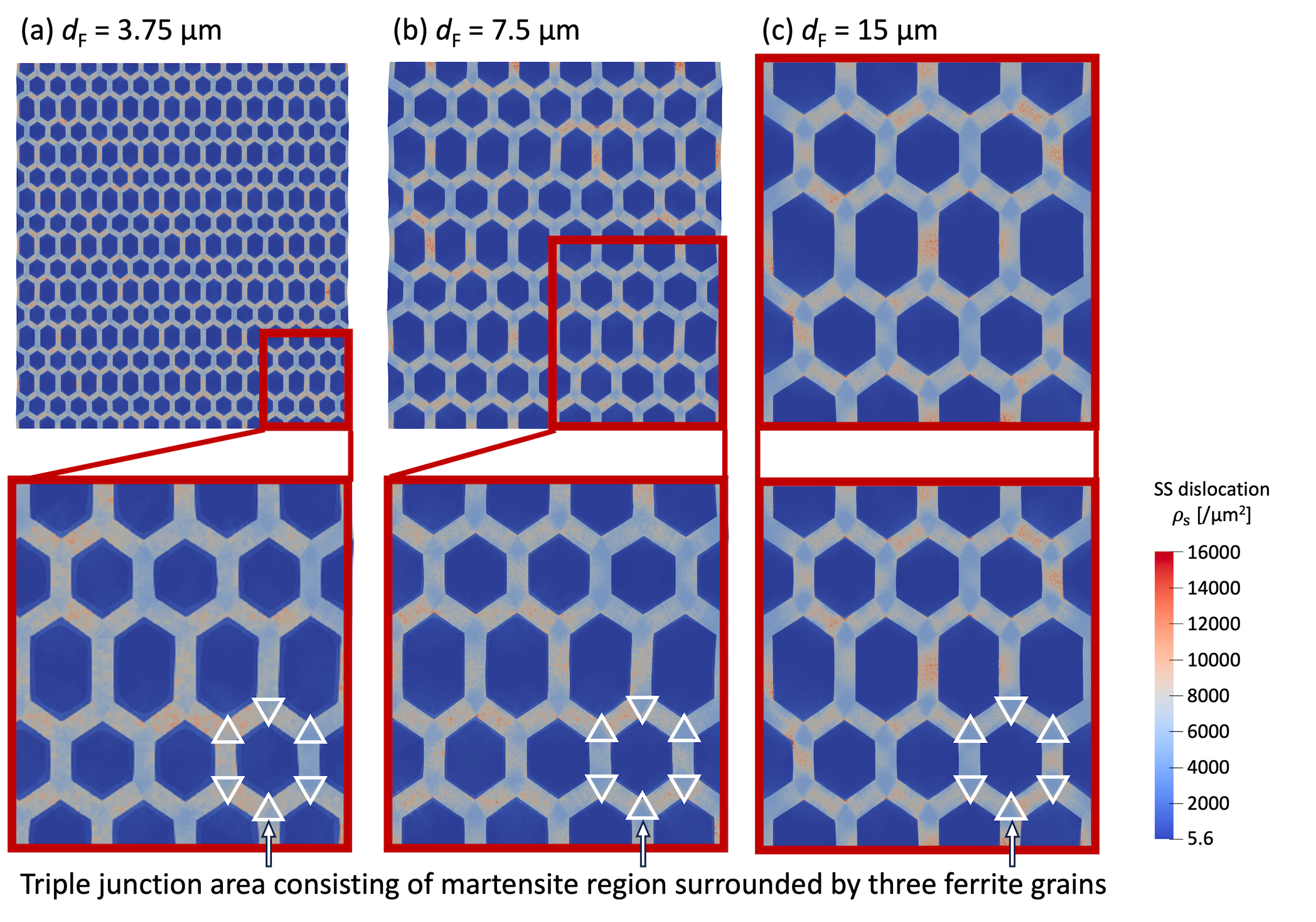}
    \caption{Distribution of SS dislocation density. (a) Grain size of ferritic phase $d_\mathrm{F}=3.75\,\si{\um}$. (b) $d_\mathrm{F}=7.5\,\si{\um}$. (c) $d_\mathrm{F}=15\,\si{\um}$. The upper figures show the distribution of SS dislocation density over the entire analytical domain. The lower figures show the results magnified so that the grain size of the ferritic phase is shown in the same size from (a) to (c). The white triangles highlight the triple junction area consisting of martensite region surrounded by the three ferrite grains.}\label{fig14}
\end{figure}

Fig.\ref{fig14}(a), (b) and (c) show the distribution of SS dislocation density for $d_\mathrm{F}=3.75$, $7.5$ and $15 \,\si{\um}$, respectively.
In Fig.\ref{fig14}, the white triangles highlight the triple junction area of martensite region surrounded by the three ferrite grains.
In this highlighted zone, the SS dislocation is lower than in other regions of the martensitic phase, and this is consistent with the trend observed in the equivalent stress distribution.
This means that the equivalent stress is low in the region of low SS dislocation density, which is also consistent with Eq.\eqref{eq9}.

\section{Conclusions}
In this study, dislocation-based crystal plasticity FE analysis is performed for DP steel microscopic specimens subject to tensile load, in which only the grain size of the ferritic phase differs, to investigate the grain size dependence of the mechanical properties and dislocation behavior of DP steel.
The findings resulting from the analysis are as follows.
\begin{itemize}
\item The martensitic phase, which is the hard phase, always shows higher stress than the ferritic phase, and its grain size dependence is also more significant than that of the ferritic phase.
This attributes to the significant increase in SS dislocation density in the martensitic phase with grain size decrease.
This trend is consistent with the experimental results.
\item The ferritic phase shows higher strain than the martensitic phase, however, this tendency becomes weaker, and the strain distribution becomes more uniform over the entire region with decreasing grain size.
This means that the contribution of the martensitic phase to deformation increases with decreasing the ferrite grain size.
As a result, the smaller the grain size, the larger the difference between the stress in the martensitic phase and that in the ferritic phase.
This trend is consistent with the experimental results.
\item GN dislocations accumulate on the ferrite side of the ferrite--martensite grain boundary, and the grain boundary occupancy per unit area increases as the ferrite grain size decreases, thus the grain size dependence of GN dislocation density is more significant in the ferritic phase than that in the martensitic phase.
\item SS dislocation density is higher in the martensitic phase than in the ferritic phase, and the grain size dependence is also more significant in the martensitic phase than that in the ferritic phase.
This attributes to the fact that as the ferrite grain size decreases, the contribution of martensitic phase to deformation is increasing, resulting in an increase in slip rate and consequently a higher SS dislocation density in the martensitic phase.
\end{itemize}

\section*{Acknowledgements}
This work was supported by CREST, Japan Science and Technology Agency (JST) Grant Number JPMJCR1994, Japan.
In addition, the authors would like to thank Professor N. Tsuji and Assistant Professor M. Park, Kyoto University for their helpful discussions and advice in carrying out this study.
% \appendix
% \section{Example Appendix Section}


\bibliographystyle{elsarticle-harv} 
\bibliography{reference.bib}


%% https://en.wikibooks.org/wiki/LaTeX/Bibliography_\mathrm{M}anagement



% \bibitem[Lamport(1994)]{lamport94}
%   \textit{\LaTeX: a document preparation system},
%   Addison Wesley, Massachusetts,

% \end{thebibliography}
\end{document}